\documentclass[reprint,superscriptaddress,amsmath,amssymb,aps,prb,showpacs]{revtex4-2}
\newcommand\HoB{HoB$_4$}
\newcommand\NdB{NdB$_4$}
\newcommand\ErB{ErB$_4$}
\newcommand\TmB{TmB$_4$}
\newcommand\TbB{TbB$_4$}
\newcommand\DyB{DyB$_4$}
\newcommand\RB{$R$B$_4$}

\usepackage{ulem}
\usepackage{graphicx,dcolumn,bm,natbib,xcolor,nicefrac}
\usepackage{soul,siunitx,color,hyperref,comment,xr}
\hypersetup{
    colorlinks=true,
    linkcolor=blue,    
    citecolor=blue,    
    filecolor=black,    
    urlcolor=blue      
    }

\begin{document}
\title{Spontaneous lattice distortion and crystal field effects in HoB$_4$}

\author{S.~Goswami}
\affiliation{Institute of Physics of the Czech Academy of Sciences, Na Slovance 2, 18200 Prague, Czech Republic}
\author{D. I.~Gorbunov}
\affiliation{Hochfeld-Magnetlabor Dresden (HLD-EMFL) and W\"urzburg-Dresden Cluster of Excellence ct.qmat, Helmholtz-Zentrum Dresden-Rossendorf, 01328 Dresden, Germany}
\author{D.~Kriegner}
\affiliation{Institute of Physics of the Czech Academy of Sciences, Na Slovance 2, 18200 Prague, Czech Republic}
\affiliation{Department of Condensed Matter Physics, Charles University, Ke Karlovu 5, 12116 Prague, Czech Republic}
\author{I.~Ishii}
\affiliation{Department of Quantum Matter, AdSE, Hiroshima University, Higashi-Hiroshima 739-8530, Japan}
\author{C. A.~Corr\^ea}
\affiliation{Institute of Physics of the Czech Academy of Sciences, Na Slovance 2, 18200 Prague, Czech Republic}
\author{T. Suzuki}
\affiliation{Department of Quantum Matter, AdSE, Hiroshima University, Higashi-Hiroshima 739-8530, Japan}
\author{D.~Brunt}
\affiliation{Department of Physics, University of Warwick, Coventry CV4 7AL, United Kingdom}
\affiliation{National Physical Laboratory, Hampton Road, Teddington TW11 0LW, United Kingdom}
\author{G.~Balakrishnan}
\affiliation{Department of Physics, University of Warwick, Coventry CV4 7AL, United Kingdom}
\author{S.~Zherlitsyn}
\affiliation{Hochfeld-Magnetlabor Dresden (HLD-EMFL), Helmholtz-Zentrum Dresden-Rossendorf, 01328 Dresden, Germany}
\author{J.~Wosnitza}
\affiliation{Hochfeld-Magnetlabor Dresden (HLD-EMFL), Helmholtz-Zentrum Dresden-Rossendorf, 01328 Dresden, Germany}
\affiliation{Institut fur Festk\"orper- und Materialphysik, TU Dresden, 01062 Dresden, Germany}
\author{O. A.~Petrenko}
\affiliation{Department of Physics, University of Warwick, Coventry CV4 7AL, United Kingdom}
\author{M. S.~Henriques}
\affiliation{Institute of Physics of the Czech Academy of Sciences, Na Slovance 2, 18200 Prague, Czech Republic}

\date{\today}

\begin{abstract}
The tetraboride \HoB\ crystallizes in a tetragonal structure (space group {\it P}4/{\it mbm}), with the Ho atoms realizing a Shastry-Sutherland lattice.
It orders antiferromagnetically at $T_{\rm N1}=7.1$~K and undergoes further magnetic transition at $T_{\rm N2}=5.7$~K.
The complex magnetic structures are attributed to competing order parameters of magnetic and quadrupolar origin with significant magnetoelastic coupling. 
Here, we investigate the response of the lattice of \HoB\  across the antiferromagnetic phase transitions by using low-temperature powder x-ray diffraction and ultrasound-velocity measurements, supported by crystal electric field (CEF) calculations. 
Below $T_{\rm N2}$, the crystal structure of \HoB\ changes to monoclinic (space group $P2_1/b$) as a macroscopic manifestation of the quadrupolar ordering.
Between 300 and 3.5~K, the total distortion amplitude is 0.46~\AA\ and the relative volume change is $3.5 \times 10^{-3}$.
This structural phase transition is compatible with the huge softening of the modulus $C_{44}$ observed around $T_{\rm N2}$ due to ferroquadrupolar order.
A lattice instability developing immediately below $T_{\rm N1}$ is seen consistently in x-ray and ultrasound data.
CEF analysis suggests a quasi-degenerated ground state for the Ho$^{3+}$ ions in this system.
\end{abstract}

\maketitle

\section{Introduction}

The 4$f$ electrons of the rare-earth elements possess spin and orbital degrees of freedom.
It is generally expected that the 4$f$ electrons are involved in crystal electric field (CEF), exchange, and multipolar interactions.
Active multipolar moments depend on the low-lying CEF levels of the 4$f$ ions and may be hard to detect.
Some microscopic techniques, e.g., neutron~\cite{Link_1998,Onimaru_2005} and resonant x-ray scattering~\cite{Paixao_2002,Okuyama_2005,Ji_2007}, have proven very useful for studies of orbital orderings.
Among macroscopic techniques, ultrasound is one of the most appropriate probes for studying orbital effects.
Elastic constants couple strain to stress in the presence of multipolar moments arising due to unquenched 4$f$ orbital moments~\cite{Yanagisawa_2003,Yanagisawa_2008,Mitsumoto_2013,Ishii_2018}.
Strong multipolar interactions frequently lead to a lowering of the crystal symmetry through electron-lattice coupling, which is reflected in a large softening of the elastic constants~\cite{Ji_2007,Watanuki_2005,Sim_2016}.

\RB\ ($R$ is a rare-earth atom) compounds offer the opportunity to study the interplay between spin, orbital, and lattice degrees of freedom.
They crystallize in a tetragonal crystal structure whose building block is a Shastry-Sutherland square lattice~\cite{Buschow_1972,Etourneau_1979,Fisk_1981}.
The $R$ ions in the \RB\ lattice feature competing nearest-diagonal and next-nearest exchange interactions and results in geometric frustration when both interactions are antiferromagnetic~\cite{Shastry_1981}.
Several magnetization plateaus were found for \ErB\ and \TmB\ as a consequence of frustration \cite{Michimura_2006,Iga_2007,Siemensmeyer_2008,Ye_2017}.
Interestingly, frustrated exchange interactions alone cannot explain the plateaus found for \TbB, and quadrupolar interactions need to be considered \cite{Yoshii_2008}.
For \DyB\, a quadrupolar ordering was found at 12.5~K that also induces a structural transition from tetragonal to monoclinic symmetry~\cite{Okuyama_2005,Ji_2007,Sim_2016}.
Quadrupolar order has also been proposed for \NdB~\cite{Yamauchi_2017}, and although no structural distortions have been experimentally detected, a symmetry-based interpretation of the magnetic structures of \NdB\ suggests that a nonmagnetic order parameter is needed to stabilize the unusual magnetic configuration~\cite{Khalyavin_2024}. 
High-resolution capacitance dilatometry experiments provided clear evidence of significant magnetoelastic coupling in \NdB ~\cite{Ohlendorf_2021,Ohlendorf_2023}.
The low-temperature structural transition proposed by Watanuki and co-authors~\cite{Watanuki_2009} has been considered as a macroscopic manifestation of quadrupolar ordering.
Latest inelastic neutron scattering data have indeed shown that the weak first-order transition at $T = 4.8$~K corresponds to the emergence of a weak magnetic-quadrupolar coupling~\cite{Metoki_2025,Yamauchi_2025} in this system.

The subject of the present study is \HoB, for which the role of the quadrupolar degrees of freedom has not yet been clarified.
Previously, it was reported that \HoB\ orders in an incommensurate antiferromagnetic (AFM) state below its N\'{e}el temperature $T_{\rm N1}=7.1$~K and exhibits a first-order transition to a noncollinear commensurate AFM phase at $T_{\rm N2}=5.7$~K~\cite{Fisk_1981,Okuyama_2008,Kim_2009,Brunt_2017}.
The magnetic transitions seem to be coupled to lattice distortions and/or structural changes~\cite{Okuyama_2008,Brunt_2017}.
Despite the relevance of the crystal structure in the framework of the quadrupole
ordering and its role concerning the competition of the different energy terms in this system, the crystal structure of \HoB\ was never thoroughly examined across the magnetic phase transitions.

Here, we focus on the elastic properties of \HoB\ in the low-temperature region.
Our detailed x-ray experiments reveal that for $T< T_{\rm N2}$ the crystal structure of \HoB\ changes to monoclinic described by the space group $P2_1/b$ as a hallmark of the quadrupolar ordering.
Our ultrasound data show a huge softening of the elastic modulus $C_{44}$ upon approaching $T_{\rm N1}$ and the coupling constants suggest ferroquadrupolar ordering.
A quasi-degenerated ground state results from our CEF analysis performed for the Ho$^{3+}$ ions in the tetragonal phase.

\section{Experimental and data analysis}

Single crystals of \HoB\ were grown and characterized as described by Brunt {\it et al}.~\cite{Brunt_2019}.
To ensure the high quality of the sample, single crystals were ground to a fine powder and used for x-ray studies.

The powder samples of \HoB\ were loaded homogeneously onto a Cu cold finger.
The powder x-ray diffraction (PXRD) study was performed in Bragg-Brentano focusing geometry and $\theta-\theta$ configuration (fixed sample) in a refurbished Siemens D500 diffractometer equipped with a closed-cycle He cryostat (101J Cryocooler from ColdEdge) at the Materials Growth \& Measurement Laboratory (MGML) in Prague~\cite{MGML}.
A Cu tube and a straight linear detector~\cite{Kriegner_2015} were used to enable the fast recording of temperature-dependent XRD data.
The sample temperature was measured on the cold finger, and thermal isolation was ensured by an evacuated sample environment and a cold shield covered by a metallized Mylar foil.

Powder diffraction patterns were collected between 300 and 3.5~K in the angular range $8^\circ \leq 2\theta \leq 135^\circ$.
Between 8 an 3.5~K, we collected diffraction patters with 0.2~K temperature steps upon cooling and heating.
For the angular region $60^\circ \leq 2\theta \leq 80^\circ$, the temperature steps were reduced and the diffraction patterns were taken every 0.1~K.
During the measurements, the temperature was allowed to stabilize after each step such that thermal equilibrium between the sample and the holder was achieved.

Whole-pattern refinements were carried out by the Le Bail and Rietveld methods implemented in the program \textsc{JANA2020}~\cite{Petricek_2023}.
Errors in our data analysis were minimized by considering the instrumental contribution to the profile width and shape of the diffracted peaks.
The numerical parameters describing the line broadening and displacement were estimated from the fitting to our experimental patterns collected at room temperature on Si and LaB$_6$ powder as standard materials supplied by the National Institute of Standards \& Technology (NIST640d and NIST660b, respectively).
Pseudo-Voigt peak-shape functions were used to describe the patterns.
The background was refined using Legendre polynomials, and the asymmetry of the diffractometer was accounted for by the Bérar-Baldinozzi correction~\cite{Berar_1993}, which proved to be best when describing the standard patterns.

Because the peaks produced by the \HoB\ sample were sharper than the ones from the NIST standards, the determination of the lattice parameters at $300 \, \mathrm{K}$ has an estimated experimental error of $1 \times 10^{-5}$~\AA.
The room-temperature lattice parameters of \HoB\ are $a=7.08576(4)$~\AA\ and $c=4.00730(2)$~\AA.

As the K$\alpha_1$ and K$\alpha_2$ wavelengths of the x-rays generated by the Cu tube of the instrument are not filtered, all peaks are composed of both components (K$\alpha_2$~/~K$\alpha_1$~=~0.48).
All the numerical results presented refer to the Cu K$\alpha_1$ component of the x-ray peaks, unless otherwise stated.

Symmetry analysis for the crystal phase transition in \HoB\ was performed using ISODISTORT~\cite{ISODISTORT,Campbell_2006} and the tools SYMMODES~\cite{Capillas_2003} and AMPLIMODES~\cite{PerezMato_2010} from the Bilbao Crystallographic Server~\cite{Bilbao}.

The field and temperature dependences of the relative sound-velocity changes in single crystal \HoB\ samples were measured using an ultrasound pulse-echo technique~\cite{Luthi_2005,Zherlitsyn_2014}.
A pair of LiNbO$_{3}$ piezo-electric transducers were glued to opposite surfaces of the sample to excite and detect acoustic waves.
We measured the longitudinal, $C_{11}$ ($\mathbf{k} \parallel \mathbf{u} \parallel$ [100]) and $C_{33}$ ($\mathbf{k} \parallel \mathbf{u} \parallel$ [001]), as well as the transverse, $C_{44}$ ($\mathbf{k} \parallel$ [100], $\mathbf{u} \parallel$ [001]) and $C_{66}$ ($\mathbf{k} \parallel$ [100], $\mathbf{u} \parallel$ [010]) modes.
Here, $\mathbf{k}$ and $\mathbf{u}$ are the wavevector and polarization of the acoustic waves, respectively.
The ultrasound measurements were supported by CEF calculations using standard methods~\cite{Luthi_2005}.

\section{Results and discussion}

\begin{figure*}[tb]
\includegraphics[width=1.5\columnwidth]{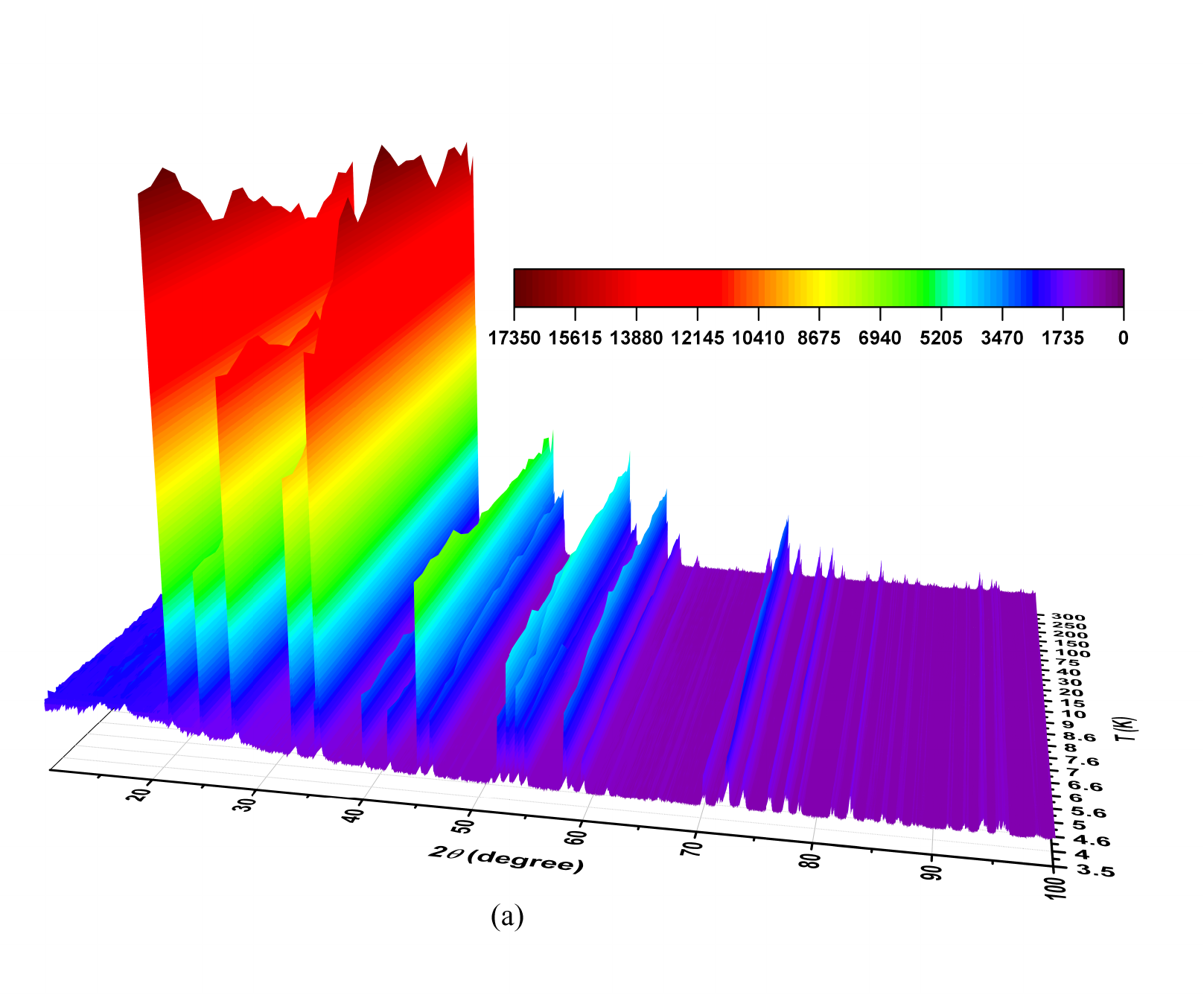}
\includegraphics[width=1.5\columnwidth]{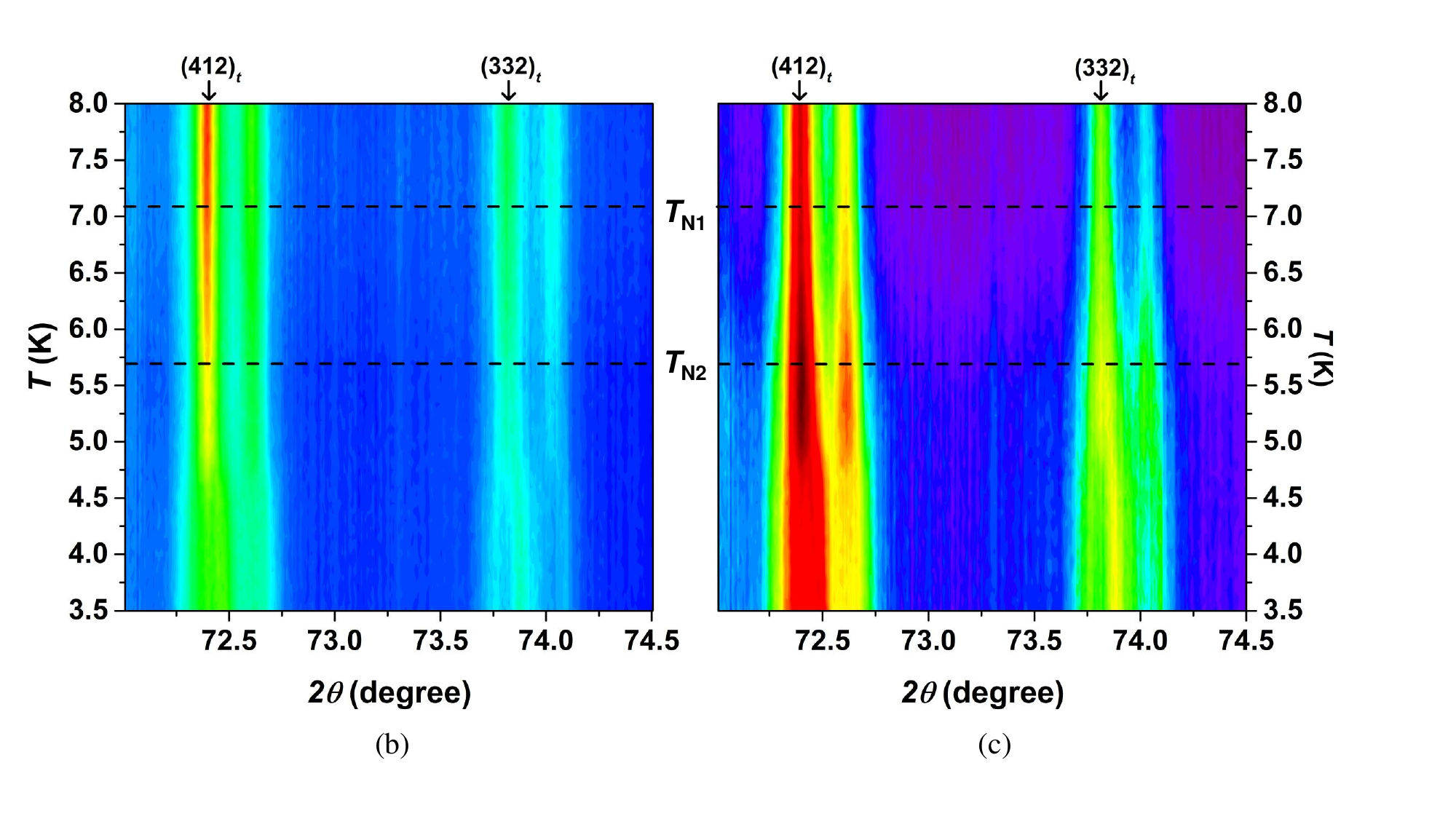}
\caption{(a) Temperature evolution of PXRD patterns of \HoB\ for $300 \, \mathrm{K} \geq T \geq 3.5 \, \mathrm{K}$.
The color scale is used to indicate the diffracted intensity (counts/s).
PXRD patterns spanning the magnetic transitions collected upon (b) cooling and (c) heating.
A clear structural distortion is seen below $T_{\rm N2}$.
The double peaks for each family of reflections in the tetragonal setting (indexed at the top) are due to the presence of Cu K$\alpha_1$ and Cu K$\alpha_2$ wavelengths.}
\label{fig:PXRD}
\end{figure*}

The PXRD patterns recorded for  \HoB\ between 300 and 3.5~K are presented in Fig.~\ref{fig:PXRD}.
No noticeable changes in the diffraction patterns of \HoB\ occur down to about 7~K, as shown in Fig.~\ref{fig:PXRD}(a).
Below this temperature, and upon cooling down to 3.5~K, the PXRD patterns show differences in intensity, with some peaks split.
The changes between 7 and 3.5~K are more noticeable for the diffraction range $50^\circ$ to $80^\circ$ [Fig.~\ref{fig:PXRD}(a)], corresponding to high-index reflections of the tetragonal symmetry (hereafter indicated by a subscript $t$)  as $(202)_t$, $(410)_t$, $(212)_t$, $(330)_t$ [$2\theta$: $50^\circ$-$55^\circ$], $(411)_t$, $(331)_t$ [$2\theta$: $58^\circ$-$62^\circ$], $(412)_t$, $(332)_t$, and $(213)_t$ [$2\theta$: $70^\circ$-$78^\circ$].

The patterns recorded during cooling and heating are presented in Figs.~\ref{fig:PXRD}(b) and \ref{fig:PXRD}(c), respectively, for the reflections $(412)_t$ at $2\theta = 72.4^\circ$ and $(332)_t$ at $2\theta \approx 73.8^\circ$ (for the Cu K$\alpha_1$ wavelength).
When cooling, the position and intensity of these peaks remain the same down to $\approx 7$~K.
Below this temperature, their intensity decreases by 44\% for $(412)_t$ and 16\% for $(332)_t$.
This tendency is maintained down to approximately 5.7~K, where peak broadening occurs that is accompanied by a distortion of the peak profiles with further temperature decrease.
For instance, the full width at half-maximum (FWHM) of the peak at $2\theta \approx 72.4^\circ$ increases from $0.11^\circ$ at 8~K to $0.25^\circ$ at 3.5~K.
The evolution of the FWHM with temperature for these reflections (and others) can be found in the Supplemental Material (Fig.~S1).
The distortion culminates in the splitting of the peak $(332)_t$ at $T \approx 5$~K.
No further structural changes are noticed down to 3.5~K.
All these observations are consistent with a reduction of the crystal tetragonal symmetry in this temperature range, as hinted in previous works~\cite{Okuyama_2008,Brunt_2017}.

Upon heating, there is hysteresis in the transition temperatures and an increase in intensity compared to the cooling process [Fig.~\ref{fig:PXRD}(c)].
That is, the peak broadening decreases at a slower rate and the split reflections of the low-temperature phase join at higher temperatures when heating.
This stems from the fact that the range of the structural phase transition is essentially dependent on the nucleation of the ‘new’ phase, which is affected by sample quality~\cite{Mnyukh_2013}.
The nucleation of the low-temperature phase begins at heterogeneous defects created during the cooling process, which are activated below $T_{\rm N1}$.
The thermalization of the sample at 3.5~K for 30~minutes leads to full transformation, stabilization, and a decrease in the number of defects in the crystals.
This is reflected in the higher intensity of the reflections and the lower background of the diffraction patterns collected upon heating [Fig.~\ref{fig:PXRD}(c)] compared to those collected during cooling [Fig.~\ref{fig:PXRD}(b)].
A very stable phase with fewer defects also explains why threshold nucleation lags upon heating, so that the tetragonal phase is reached at higher temperatures and a wide hysteresis is found for samples of high quality.

\begin{figure}[tb]
\includegraphics[width=0.99\columnwidth]{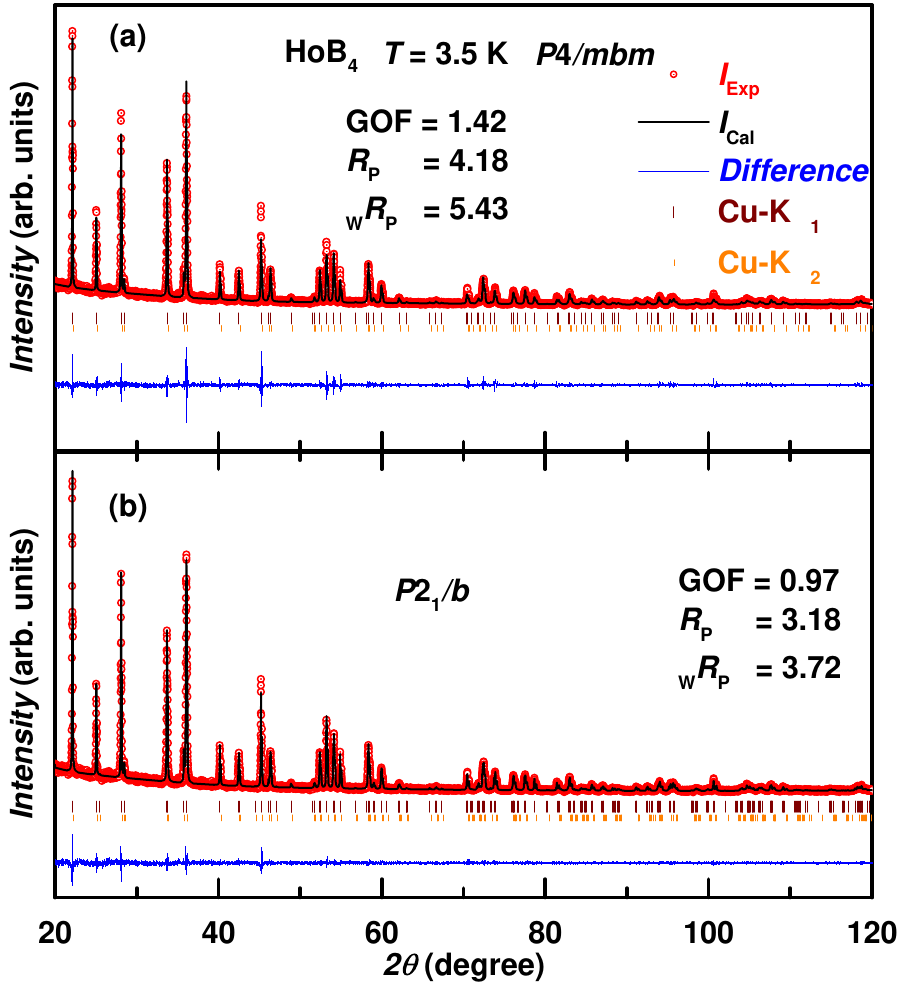}
\caption{Le Bail fitting to the PXRD pattern at 3.5~K treated with (a) the tetragonal space group $P4/mbm$ and (b) with the monoclinic space group $P2_1/b$.}
\label{fig:PXRD_LeBail}
\end{figure}

The PXRD patterns at 3.5~K were initially treated considering the reported tetragonal crystal symmetry (space group $P4/mbm$)~\cite{Etourneau_1979} using the program JANA2020~\cite{Petricek_2023}.
First, Le Bail analysis (profile fitting accounting only for the generated peak positions according to the space group of the compound~\cite{LeBail_2005}) was performed.
The difference line between the calculated and experimental patterns is shown in Fig.~\ref{fig:PXRD_LeBail}(a).
The fit with the tetragonal symmetry does not describe properly several peaks, especially for $2\theta > 40^\circ$.
The calculated intensities for the Cu K$\alpha_2$ component are not well resolved by this space group, which is more evident for some of the reflections such as $(210)_t$ at $2\theta = 28^\circ$ and $(310)_t$ at $2\theta = 40^\circ$.
These results are not unexpected as the compound no longer has tetragonal symmetry, and it is better described by another space group of symmetry lower than $P4/mbm$.
Prior research on this material~\cite{Okuyama_2008,Brunt_2017} also noticed possible lattice distortions and a phase transition.

The structural phase transition of \HoB\ from $P4/mbm$ to any of its possible subgroups was checked with the help of JANA2020 and the SUBGROUPS tool from the Bilbao Crystallographic Server~\cite{Ivantchev_2000}.
The best match between experimental and calculated patterns at 3.5~K was obtained for the monoclinic space group $P2_1/b$, as shown in Fig.~\ref{fig:PXRD_LeBail}(b).
The goodness of fit (GOF) and the quality of other fit measures are significantly improved with this space group, as can be seen in Fig.~\ref{fig:PXRD_LeBail}.
Moreover, the reflection conditions for both space groups (tetragonal and monoclinic) are very similar, and the Bragg peaks are generated at close positions, making it difficult to assign \textit{a priori} a certain peak to a specific crystal symmetry.
A few higher-order reflections differentiating the two symmetries, such as $(104)_m$ and $(304)_m$ (allowed only for the monoclinic symmetry) appearing at a much higher scattering angle ($2\theta > 100^\circ$), have a negligible intensity.
Thus, Le Bail analysis reveals that the crystal lattice of \HoB\ undergoes a symmetry reduction from tetragonal to monoclinic at $T \approx T_{\rm N2}$. 
Further, the splitting observed for some reflections shown in Fig.~\ref{fig:PXRD} is also consistent with the tetragonal to monoclinic symmetry change.
For example, the tetragonal $(332)_t$ reflection at $2\theta \approx 73.8^\circ$ has multiplicity 8.
Below $T \approx5$~K, it splits into the non-equivalent monoclinic reflections $(3\text{-}32)_m$ and $(332)_m$ (both with multiplicity 4) at the positions $2\theta \approx 73.84^\circ$ and $2\theta \approx 73.9^\circ$, respectively.

The crystallographic group-subgroup relations between the space groups $P4/mbm$ and $P2_1/b$ suggest that the structure of \HoB\ should transform through the orthorhombic space group \textit{Pbam}.
Nevertheless, during this work no clear orthorhombic distortion was detected despite the attempts to fit the patterns taken for $T_{\rm N1} \geq T \geq T_{\rm N2}$ to the orthorhombic space group. 
Additional detailed inspection of reflections that could be used to track a tetragonal-orthorhombic transition (as 200, 202/022 or even 412) was performed (Fig.~S2 Supplemental Material). The FWHM of the peaks generally increases below $T_{\rm N1}$ compared to room temperature (Fig.~S1 Supplemental Material), but it is not significant to indicate changes of the lattice parameters and loss of tetragonal symmetry.  
 
However, one has to keep in mind the limitations of the conventional powder diffraction technique and of the procedures applied for data treatment.
The main factors that may prevent the observation of an intermediate structural distortion include: 
(i) the averaging of equivalent reflections into a single peak due to the same structure factor causes loss of information, (ii) algorithms to extract/establish the peak positions or indexing overlook some of the experimental data, (iii) the instrumental resolution is not sufficient to reveal peak splitting, (iv) the transition is too weak to be observed by PXRD.
In addition, the temperature span for the structural distortion is very narrow in the present case.

\begin{figure*}[tb]
\includegraphics[width=1.60\columnwidth]{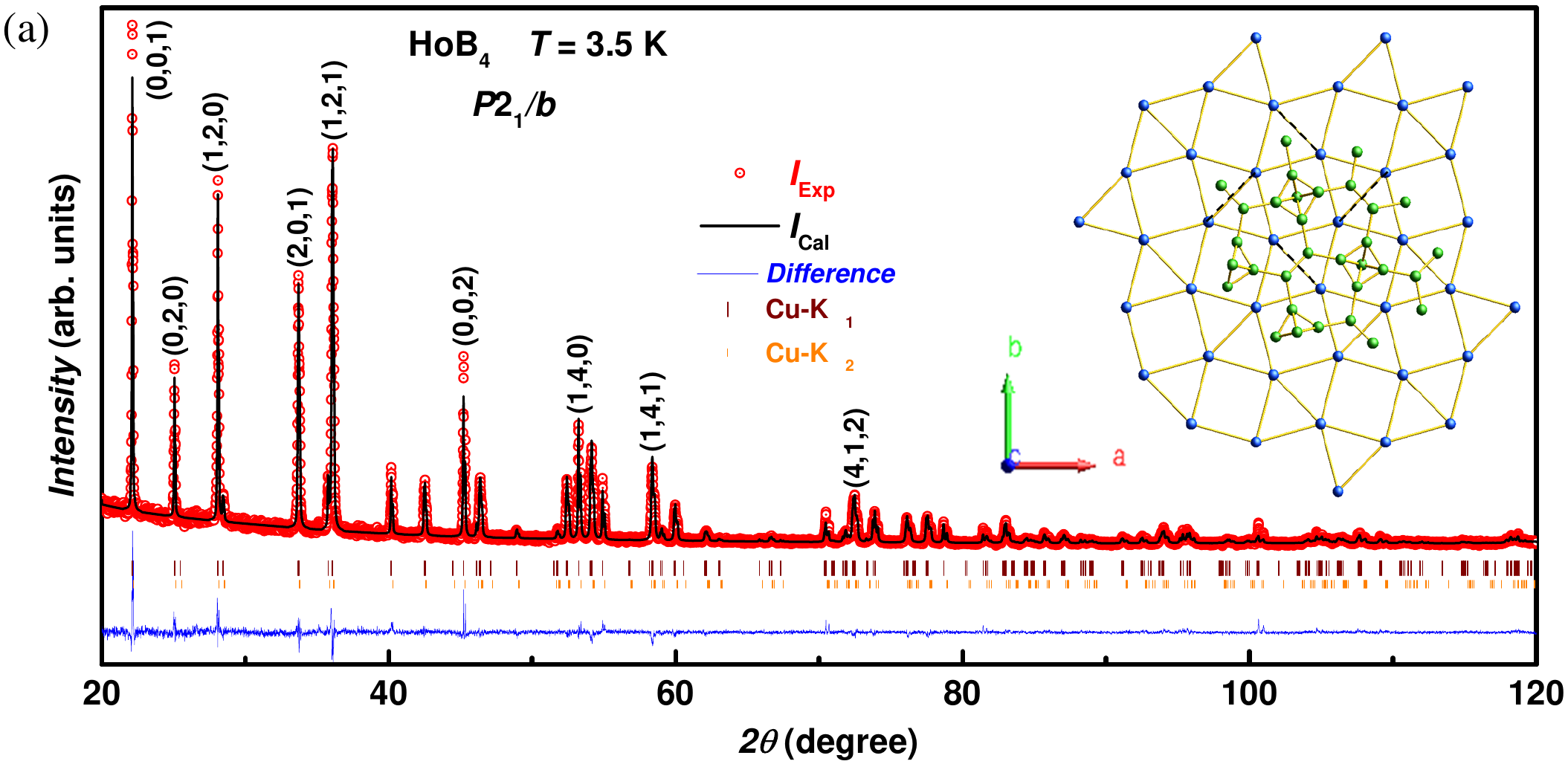}
\includegraphics[width=1.60\columnwidth]{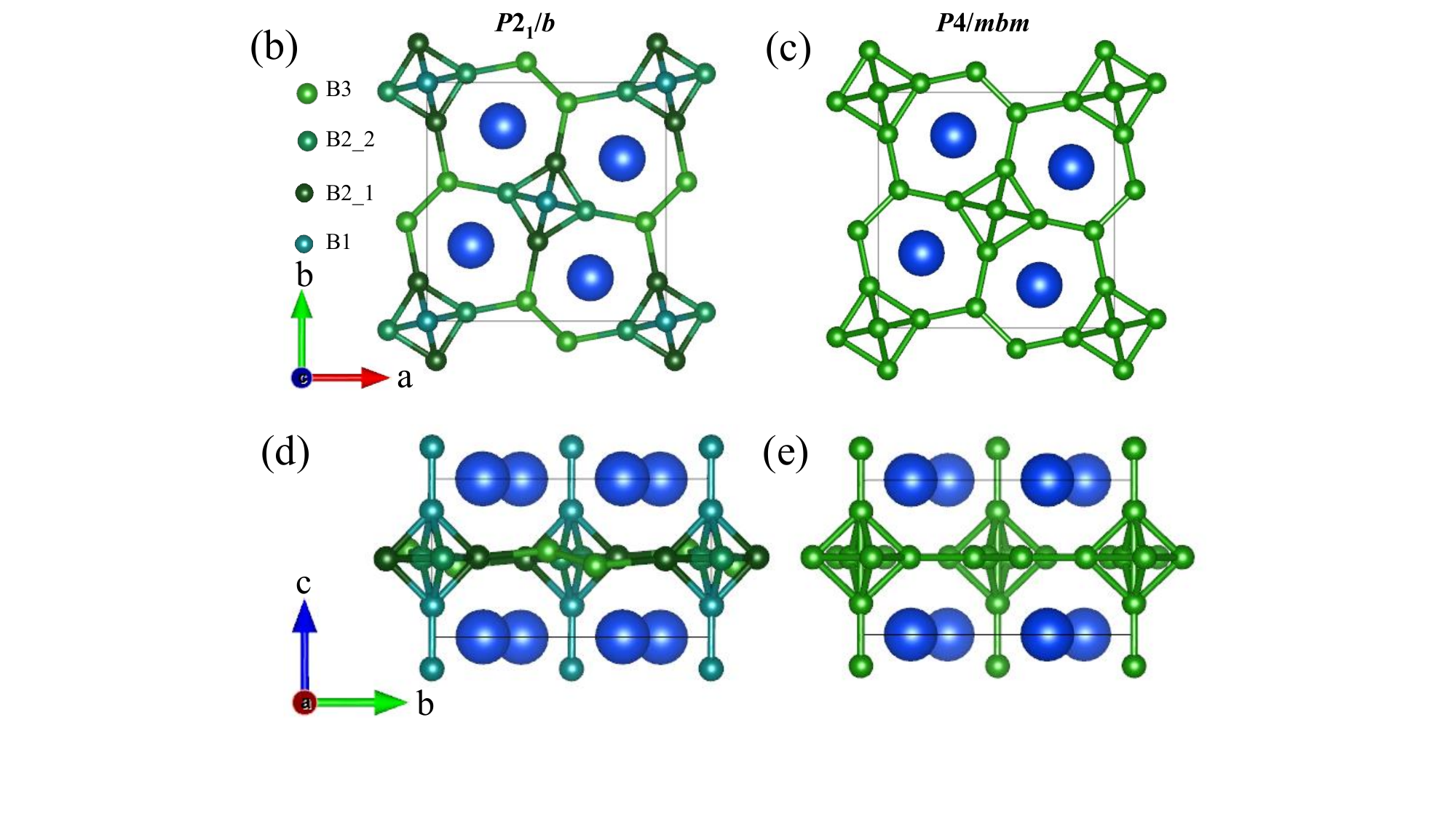}
\caption{(a) Rietveld-refined PXRD pattern of \HoB\ at 3.5~K using the space group $P2_1/b$.
Corresponding monoclinic crystal structure of \HoB\ at 3.5~K: view
along (b) the $c$~axis and (d) the $a$~axis.
The same projections for the tetragonal structure for comparison are shown in (c) and (e).
Ho atoms are depicted in blue, whereas the B atoms are displayed
in green. Labels for the different B positions in the monoclinic unit cell in (b) refer to Table I.
Black dashed lines in the inset of (a) show that the Ho sublattice in the monoclinic phase preserves the geometry of a Shastry-Sutherland lattice.}
\label{fig:Rietveld}
\end{figure*}

Rietveld refinement~\cite{Rietveld_1967,Rietveld_1969} of the PXRD pattern of \HoB\ at 3.5~K was performed, and the result is shown in Fig.~\ref{fig:Rietveld}(a).
The refinement converged successfully to a good fit in the space group $P2_1/b$.
The fitting factors, lattice parameters, and the refined position of each atom can be found in Table~\ref{tab:PXRD}. It is interesting to note that the monoclinic angle refined for the low-temperature structure deviates only by $0.127^\circ$ from the tetragonal $90^\circ$.
A similar value of the monoclinic angle was estimated by Okuyama \textit{et al.}~\cite{Okuyama_2008} for this compound, while a slightly larger deviation was found for the Dy counterpart from high-resolution powder neutron and synchrotron x-ray diffraction~\cite{Sim_2016}.

In the monoclinic phase of \HoB, the independent Ho and B (1Ho + 4B) in the unit cell are at a general position (4e: $x,y,z$) when compared to the tetragonal symmetry, for which four independent atoms (1Ho + 3B) were kept at special crystallographic positions.
The coordinates of the Ho atom do not change much compared to their values in the $P4/mbm$ unit cell. The position of the atoms along the main crystallographic directions can be compared for monoclinic and tetragonal unit cells in Figs.~\ref{fig:Rietveld}(b) and \ref{fig:Rietveld}(c).
It should be noted that the nearest Ho-Ho distance in the monoclinic symmetry is very close to the one in the tetragonal unit cell ($\approx 3.66$~\AA).
The $(ab)$ projection displayed in Figs.~\ref{fig:Rietveld}(b) and \ref{fig:Rietveld}(c) seems to indicate that the B sublattice is similar for both symmetries.
However, the projection $(bc)$ reveals the distortion of the B chains along the $b$~axis in the monoclinic unit cell [Fig.~\ref{fig:Rietveld}(d)] when compared to the tetragonal one [Fig.~\ref{fig:Rietveld}(e)].
In fact, the total distortion amplitude (square root of the sum of the square of all atomic displacements within a primitive unit cell of the tetragonal structure) calculated for this transition is about 0.46~\AA\, with a maximum atomic displacement of approximately 0.19~\AA\ found for the B3 atoms in the distorted chains [Fig.~\ref{fig:Rietveld}(d)]. 

At the phase transition, the B atoms at the tetragonal $8j$ $(x,y,0.5)$ position (B2) split into two monoclinic $4e$ positions, designated as B2\_1 and B2\_2.
This splitting of the B2 atoms changes the bond distances between all atoms.
In the $P4/mbm$ space group, the distance between any two adjacent B2 atoms in the $ab$~plane of the octahedra is 1.563(19)~\AA, and that between the oppositely placed B2 atoms is 2.210(18)~\AA.
In contrast, for the $P2_1/b$ space group, the adjacent atoms in the octahedra in the $ab$ plane are B1 and B2 being 1.681(7)~\AA\ apart.
The distance between opposite B2 atoms is 2.372(6)~\AA, while the distance between the B1 atoms is 2.382(7)~\AA.
Because of this splitting of the $8j$ position, the distance between B3 atoms also changes from 2.378(10)~\AA\ to 2.398(7)~\AA\ in monoclinic symmetry.
A list of interatomic distances in the monoclinic crystal structure can be found in Table~\ref{tab:InteratomicDistances}.

\begingroup
\squeezetable
\centering
\begin{table}[tb]
\caption{\label{tab:PXRD} Refined structural parameters and fitting factors for PXRD data of \HoB\ at $T=3.5$~K using space group $P2_1/b$.}
\begin{ruledtabular}
\begin{tabular}{ccccc}
\multicolumn{5}{l}{GOF = 1.13; $R_p = 3.62\%$; $wR_p = 4.36\%$} \\ 
\multicolumn{5}{l}{Lattice parameters:} \\
\multicolumn{5}{l}{$a = 7.07899(8)$~\AA, $b = 7.07916(8)$~\AA, $c = 4.00126(4)$~\AA} \\
\multicolumn{5}{l}{$\alpha = 90.1269(7)^\circ$, $\beta = 90^\circ$, $\gamma = 90^\circ$; Volume = $200.516(4)$~\AA$^3$} \\ \hline 
Atom & Wyckoff & \textbf{x} & \textbf{y} & \textbf{z} \\
    & Position &  &  &  \\
\hline
Ho & 4e & 0.1833(3) & 0.3180(3) & -0.0009(4) \\
B1 & 4e & 0.5010(4) & 0.5010(2) & 0.2003(13) \\
B2\_1 & 4e & 0.0373(5) & 0.8360(7) & 0.495(4) \\
B2\_2 & 4e & 0.3367(6) & 0.5373(8) & 0.500(7) \\
B3 & 4e & 0.0855(12) & 0.5848(12) & 0.453(3) \\
\end{tabular}
\end{ruledtabular}
\end{table}
\endgroup

\begingroup
\squeezetable
\begin{table}[ht]
\caption{\label{tab:InteratomicDistances} Refined interatomic distances for the atoms of \HoB\ at $T=3.5$~K in the monoclinic $P2_1/b$ space group. The number between the atomic labels indicates the number of bonds at the specified distance.}
\begin{ruledtabular}
\begin{tabular}{cc|cc}
Atoms & Distance (\AA) & Atoms & Distance (\AA) \\ \hline
Ho 2 Ho        & 3.668(3)                & B2\_1 1 B1     & 1.695(18)               \\
Ho 1 Ho        & 3.656(3)                & B2\_1 1 B1     & 1.685(18)               \\
Ho 2 Ho        & 3.664(3)                & B2\_1 1 B3     & 1.818(10)               \\
B1 1 B2\_1     & 1.695(18)               & B2\_2 1 B2\_1  & 1.680(6)                \\
B1 1 B2\_1     & 1.685(18)               & B2\_2 1 B2\_1  & 1.681(7)                \\
B1 2 B2\_2     & 1.69(3)                 & B2\_2 2 B1     & 1.69(3)                 \\
B1 1 B1        & 1.603(7)                & B2\_2 1 B3     & 1.819(10)               \\
B2\_1 1 B2\_2  & 1.680(6)                & B3 1 B2\_1     & 1.818(10)               \\
B2\_1 1 B2\_2  & 1.681(7)                & B3 1 B2\_2     & 1.819(10)               \\
B3 1 B3        & 1.746(12)               &                 &                         \\
\end{tabular}
\end{ruledtabular}
\end{table}
\endgroup

In the monoclinic symmetry, the octahedra formed by the B atoms are elongated compared to the tetragonal structure.
The distances between the Ho atoms in their monoclinic positions still satisfy the criteria for the Shastry-Sutherland lattice [inset of Fig.~\ref{fig:Rietveld}(a)].

\begin{figure}[tb]
\includegraphics[width=0.95\columnwidth]{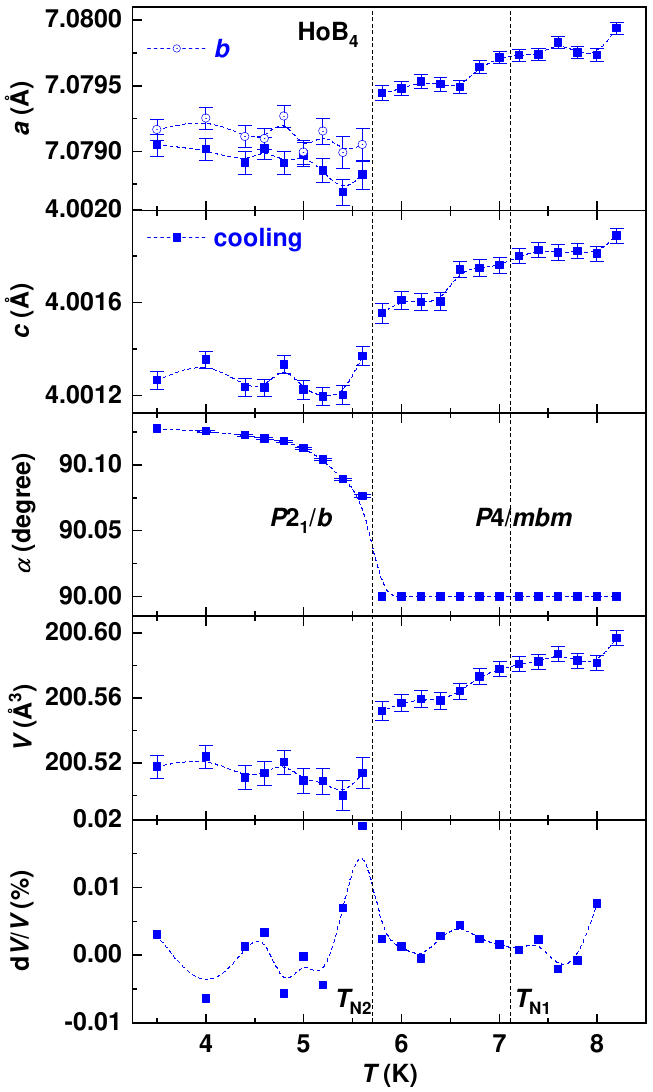}
\caption{Temperature dependence of lattice parameters, angle, unit-cell volume, and volume change between 8 and 3.5~K in \HoB\ upon cooling.
The broken lines are guides to the eye.}
\label{fig:Tvariation}
\end{figure}

Further, the PXRD data of \HoB\ collected between 3.5 and 8~K during cooling and heating were analyzed to extract the lattice parameters and the unit-cell volume following both crystal symmetries, i.e., with space group $P4/mbm$ and $P2_1/b$.
The results of the full-pattern Rietveld fittings of the data taken during cooling are plotted in Fig.~\ref{fig:Tvariation}. 

\begin{figure}[tb]
\includegraphics[width=0.99\columnwidth]{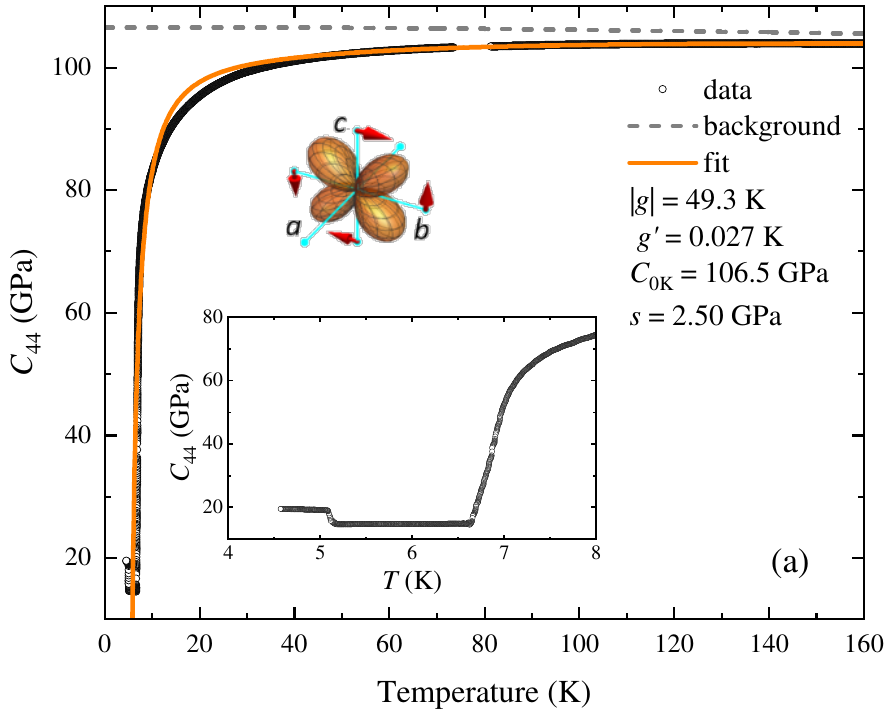}
\includegraphics[width=0.99\columnwidth]{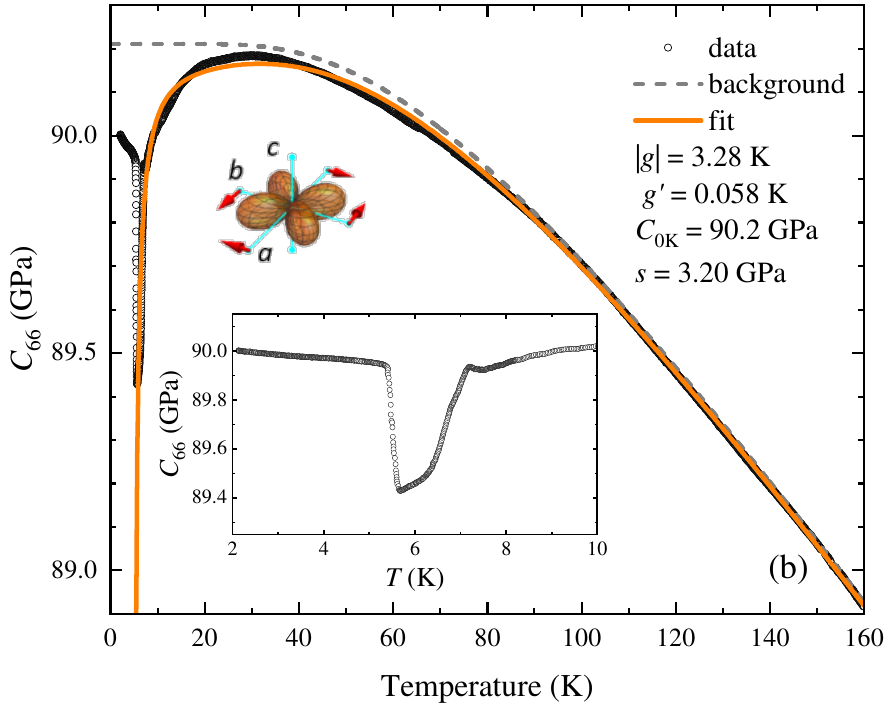}
\caption{Temperature variation of the elastic modules in \HoB\ single crystal in zero applied field for two transverse acoustic modes, (a) $C_{44}$ and (b) $C_{66}$.
Black symbols depict the experimental data, the orange lines show the calculations, while the dashed gray lines represent background contributions~\cite{Varshni_1970}.
The insets focus on the low-temperature behavior.
See main text for further explanations of the notations used.}
\label{fig:C44_C66}
\end{figure}

The lattice parameters and unit-cell volume show a smooth change in the tetragonal structure upon cooling.
All lattice parameters decrease with temperature until $T \approx T_{\rm N2}$, where a sharp step occurs, indicating the transformation to the monoclinic phase.
Here, the volume change reaches its maximum of $\approx 0.02\%$.
This behavior is expected for a first-order phase transition.
Below $T_{\rm N2}$, there is a region where the lattice parameters fluctuate, especially $a$ and $b$, before reaching the final stable values at 3.5~K.
This is also reflected by the trend of the monoclinic angle and volume changes [and in the ultrasound measurements presented later (Fig.~\ref{fig:C44_C66})].
In the interval $T_{\rm N1} \geq T \geq T_{\rm N2}$, there is a small step around 6.6~K, visible in the temperature variation of the lattice parameters $a$ and $c$ in the cooling data, which is also seen as hysteresis between heating and cooling (Fig. S3, Supplemental Material). 
This coincides with the region where the FWHM increases (Fig. S1, Supplemental Material).
These features suggest that the onset of the distortion/instabilities of the tetragonal phase in \HoB\ might occur above $T_{\rm N2}$.
This effect is more pronounced for the $a$ parameter, which could hint at an orthorhombic distortion.
Attempts to refine the patterns with an orthorhombic space group were made for the data between 6 and 7~K (Fig. S2, Supplemental Material). 
Although the refinement in the $Pbam$ space group converges, it does not reproduce well the peak asymmetry of the profile. 
However, this effect can be accounted for by imposing the condition  \textit{a} = \textit{b} during the refinement cycles. That is, the structure is tetragonal.

The overall change in the lattice parameters of \HoB\ between 3.5 and 8~K is not large. In fact, the relative volume change is $3.5 \times 10^{-3}$. This agrees well with previous studies of thermal lattice properties of rare-earth tetraborides, including this compound~\cite{Novikov_2015}.

While studying the magnetic structures of \HoB\ by neutron diffraction, Brunt \textit{et al.} noticed that below $T_{\rm N2}$ the reflections (140), (330), and their Friedel pairs, exhibited unphysical negative magnetic diffraction intensities upon cooling followed by an increase in the magnetic field~\cite{Brunt_2017}.
The negative magnetic intensities were the result of the subtraction of the nuclear diffraction intensities from the paramagnetic phase.
This unphysical scenario was resolved by taking into account a decrease in the nuclear contribution of these peaks due to changes in the atomic positions within the unit cell~\cite{Brunt_2017}.
 
In the present work, the same rationale is followed but using the monoclinic space group $P2_1/b$.
To distort the unit cell, the fractional coordinates ($x,y,z$) of Ho and B atoms are displaced by adding the values $x_1$, $y_1$, and $z_1$ respectively such that $x^\prime = x + x_1 = x + nx$, $y^\prime = y + y_1 = y + ny$, and $z^\prime = z + z_1 = z + nz$, where $n$ varies within $\pm 0.1$, as $\vert n \vert > 0.1$ would destroy the interatomic bonds between the atoms.
Here, the same atoms that were considered in the earlier work~\cite{Brunt_2017} are displaced from the refined structural positions.
The variation of the calculated structure factor or intensity versus the parameter $n$ is shown in Fig.~\ref{fig:StructureFactor}.
The trend found here reproduces the one obtained in earlier work (see Supplemental Material of~\cite{Brunt_2017}).
This means that the monoclinic model is correct, and that Brunt \textit{et al.} were right in pointing out that a lattice distortion involving displacement of the Ho and B2 sublattices could explain the apparent ‘negative’ nuclear intensity in their neutron experiments on \HoB~\cite{Brunt_2017}.

\begin{figure}[tb]
\includegraphics[width=0.99\columnwidth]{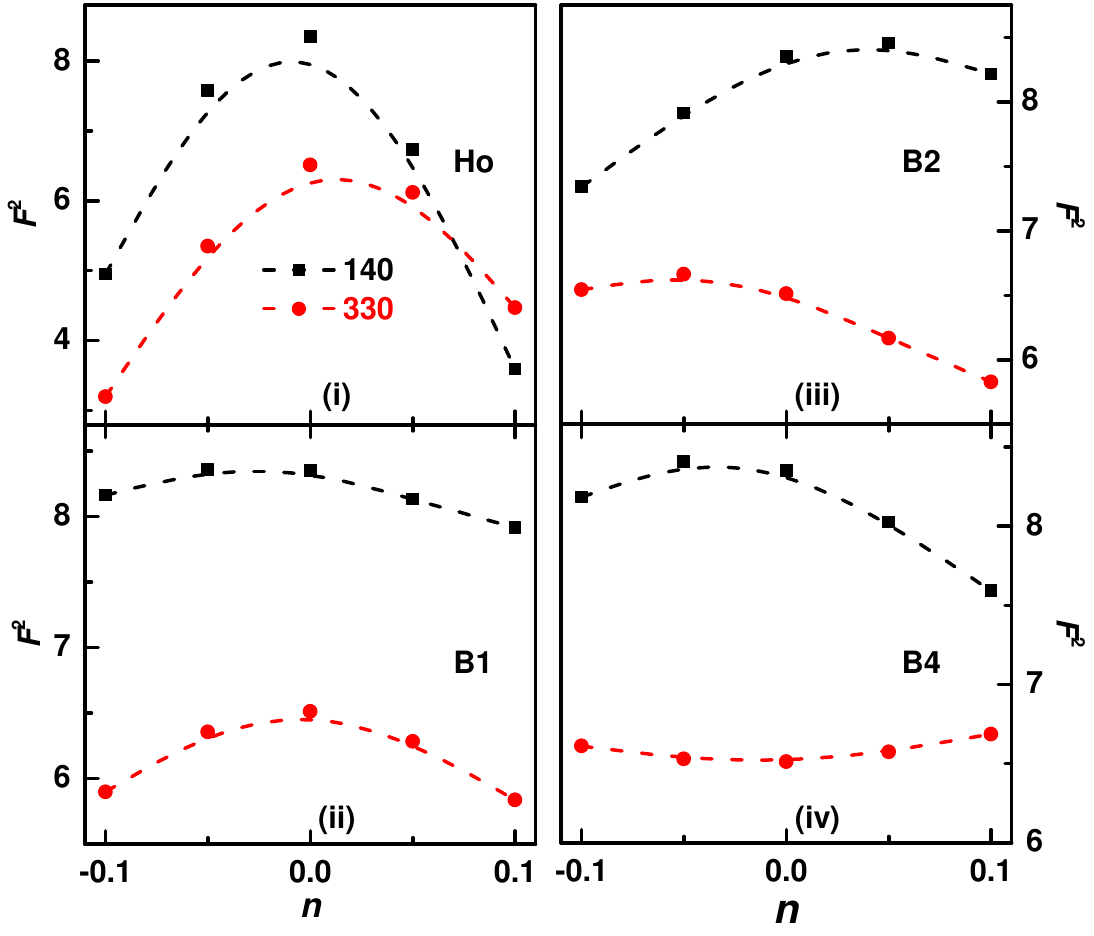}
\caption{Variation of the calculated crystal structure factor squared ($F^2$) for the reflections (140) and (330) versus the displacement parameter $n$ (defined in the text), when the atom specified (in the figure) is displaced in the monoclinic unit cell. The dashed lines are guides to the eye.}
\label{fig:StructureFactor}
\end{figure}

The calculated nuclear and magnetic structure factors for (140) and (330) are presented in Table~\ref{tab:StructureFactors} for the tetragonal and monoclinic space groups.
The magnetic structure factors for the commensurate AFM structure of \HoB\ ($T < T_{\rm N2}$) were obtained from powder simulations performed in JANA2020, considering the experimental conditions described by~\cite{Brunt_2017}.
The procedure deviates from the representation analysis embedded in the magnetic option of JANA2020~\cite{SousaHenriques_2024}.
The program calculates the possible magnetic space groups (MSGs) consistent with the parent space group $P4/mbm$ and the magnetic propagation vector.
From the list of possible MSGs, only $P2_1/c^\prime$ was considered because it corresponds to the description of the magnetic structure of \HoB\ below $T_{\rm N2}$ (in zero field) given by Okuyama \textit{et al.}~\cite{Okuyama_2008}. 
The moments are arranged 2-in-2-out in-plane with a small out-of-plane component. Both components are AFM (the representational analysis is discussed in a subsequent section). The magnetic contribution to the reflections (140) and (330) for the model is given in Table~\ref{tab:StructureFactors}. The results show that for both reflections, the diffraction intensity is considerably lower in the monoclinic symmetry, in agreement with the previous observations~\cite{Brunt_2017}.
Further, despite the contribution of the magnetic intensity to the total intensity, this sum is still lower in the monoclinic phase than the nuclear term alone in the tetragonal phase.

\begingroup
\squeezetable
\begin{table}[tb]
\caption{\label{tab:StructureFactors} Calculated nuclear $\vert F_\text{c} \vert$ and magnetic $\vert F_\text{Mc} \vert$ structure factors for the indicated reflections in the tetragonal and monoclinic symmetries.
The magnetic contribution was simulated for the monoclinic MSG.}
\begin{ruledtabular}
\begin{tabular}{ccccc}
Symmetry & Reflection & Multiplicity & $\vert F_\text{c} \vert$ & $\vert F_\text{Mc} \vert$ \\ \hline
$P4/mbm$ & (330) & 4 & 115.94 & -- \\
         & (140) & 8 & 123.77 & -- \\ \hline
$P2_1/c^\prime$ & (330) & 4 & 103.27 & 8.67 \\
         & (140) & 4 & 109.36 & 16.40 \\
\end{tabular}
\end{ruledtabular}
\end{table}
\endgroup

The analysis presented so far suggests that the transformation from tetragonal to monoclinic symmetry in \HoB\ takes place below $T_{\rm N1}$.
The crystal structure below $T_{\rm N2}$ is best described by the space group $P2_1/b$.
Such symmetry reduction for this compound was previously suggested by Okuyama \textit{et al.}~\cite{Okuyama_2008}.
Further, they concluded that the monoclinic distortion occurs around $T_{\rm N2}$ and is induced by the ordering of the Ho quadrupolar moments through strong spin-orbit coupling.
Thus, the full structural changes in \HoB\ presented here confirm the expected spontaneous symmetry breaking associated with the quadrupolar order.

A similar structural transition is well documented for \DyB~\cite{Sim_2016}.
Sim {\it et al.} demonstrated that the low-temperature monoclinic distortion (space group $P2_1/a$) is coupled to a quadrupole-moment ordering within the low-temperature magnetic phase, with the structure having magnetic moment components along [100] and [001]~\cite{Sim_2016}.
The structural transition is a lifting mechanism of the frustration in the Shastry-Sutherland lattice.
The presence of quadrupolar order has also been suggested previously for \NdB\ ~\cite{Yamauchi_2017}, but no clear structural signature was found.
Recently, Khalyavin \textit{et al.}~\cite{Khalyavin_2024} verified that the ground state magnetic structure of \NdB\ consists of two distinct components stabilized by a nonmagnetic order parameter implying a monoclinic lattice distortion as given by symmetry-based analysis.
The authors observed an anomaly in the lattice parameters as a function of temperature, but no further evidence of a possible monoclinic strain in the system was detected.
The representational analysis performed in that study has revealed that the nonmagnetic distortion required at the ground state of \NdB\ is given by the irrep GM5$^+$ (0, $a$).
Their analysis has shown that this distortion allows the symmetry lowering from tetragonal to monoclinic and satisfies the energy invariance with the magnetic and quadrupolar orders.
In \HoB, the tetragonal to monoclinic symmetry reduction requires the same primary order parameter GM5$^+$ (0, $a$).

Representation theory together with profile simulations were performed for \HoB\ by Okuyama \textit{et al.}~\cite{Okuyama_2008} to describe the magnetic ground state.
They attributed ad hoc letters to represent the possible magnetic irreps and concluded that the magnetic structure below $T_{\rm N2}$ is described by an in-plane 2-in-2-out moment component and an orthogonal component transformed from the paramagnetic $P4/mbm$ space group according to a one-dimensional (1D) magnetic irrep and a two-dimensional (2D) one, respectively.
These irreps correspond to mGM2$^-$ (1D) and mGM5$^-$ (2D) in CDML (Cracknell, Davies, Miller, and Love) notation~\cite{Cracknell_1979,Miller_1967}, respectively. Using this information and the tool \textsc{K-SUBGROUPSMAG} of the Bilbao Crystallographic Server~\cite{Bilbao}, we inferred that the magnetic space group describing the magnetic structure of \HoB\ at zero field is most probably $P2_1/c^\prime$.

Given the crucial role played by spin-orbit coupling in \HoB, we probed the elastic properties of this compound by measuring the change in the sound velocity, $\Delta v/v$, with temperature for different longitudinal and transverse acoustic modes. The sound velocity, $v$, is related to the concomitant elastic constant $C_{ij} = \rho v^2$, where $\rho$ is the mass density of the material. Figure~\ref{fig:C44_C66} shows the temperature dependence of the two elastic moduli, $C_{44}$ and $C_{66}$, obtained from our sound-velocity measurements.
While the temperature dependence of the $C_{66}$ modulus looks typical for magnetic materials with relatively low magnetic ordering transitions, the  change of the $C_{44}$ modulus is unexpectedly large (it corresponds to almost a complete softening of this transverse acoustic mode).

In the analysis of the ultrasound results, we used a Hamiltonian which includes three terms -- the CEF contribution, strain-quadrupole interaction, and quadrupole-quadrupole interaction~\cite{Hutchings_1964}.
The values of $g$ and $g^\prime$, the strain-quadrupole and quadrupole-quadrupole coupling constants, respectively, obtained from describing $C_{44}(T)$ and $C_{66}(T)$ are given in Fig.~\ref{fig:C44_C66}.
The quadrupole-quadrupole coupling constant of $C_{44}$ is positive, suggesting ferroquadrupolar ordering.
We used the Varshni equation, $C_0 = C_{\rm 0K} - \frac{s}{\exp(t/T)-1}$~\cite{Varshni_1970}, to model the background stiffness with the parameters $C_{\rm 0K}$ and $s$ as given in Fig.~\ref{fig:C44_C66} and $t=200$~K.

\begingroup
\squeezetable
\begin{table}[tb]
\caption{\label{tab:LowESinglets} Calculated energies (in kelvin) of the 9 lower-energy singlets of the total 17 states (5 $\Gamma_1$, 4 $\Gamma_2$, 4 $\Gamma_3$, 4 $\Gamma_4$) for the $C_{2h}$ orthorhombic environment of the Ho\textsuperscript{3+} ions. $J = 8$, $L = 6$, $S = 2$.}
\begin{ruledtabular}
\begin{tabular}{ccccccccc}
$\Gamma_4$ & $\Gamma_2$ & $\Gamma_1$ & $\Gamma_3$ & $\Gamma_1$ & $\Gamma_2$ & $\Gamma_3$ & $\Gamma_4$ & $\Gamma_4$ \\ \hline
0 & 6.3 & 16.1 & 33.3 & 59 & 108 & 114 & 196 & 258 \\
\end{tabular}
\end{ruledtabular}
\end{table}
\endgroup

\begingroup
\squeezetable
\begin{table}[tb]
\caption{\label{tab:CEFParameters} Calculated CEF Stevens parameters $B_{kq}$ (in kelvin) for the $C_{2h}$ orthorhombic environment of the Ho\textsuperscript{3+} ions ($4f^{10}$ configuration).}
\begin{ruledtabular}
\begin{tabular}{ccccccccc}
$B_{02}$ & $B_{22}$ & $B_{04}$ & $B_{24}$ & $B_{44}$ & $B_{06}$ & $B_{26}$ & $B_{46}$ & $B_{66}$ \\ \hline
0.61 & 0.03 & 0.0199 & 0.0128 & 0.109 & 0.000101 & 0.000598 & -0.00124 & 0.00005 \\
\end{tabular}
\end{ruledtabular}
\end{table}
\endgroup

The CEF analysis performed for the orthorhombic $C_{2h}$ environment of the Ho$^{3+}$ ions shows that all states are non-magnetic singlets, with the lowest energy $\Gamma_4$ and $\Gamma_2$ singlets forming a quasi-degenerate ground state.
There are 17 singlets in total; however, as the high-energy states (above 300~K) are not considered important in the context of this study, the estimated energy levels are not particularly accurate.
Therefore, Table~\ref{tab:LowESinglets} lists only the 9 low-energy states.
In performing the CEF analysis, we were guided not only by the temperature dependences of the sound velocity and (previously published) magnetization and magnetic susceptibility data~\cite{Brunt_2018}, but also by the results of inelastic neutron scattering (INS) performed on a powder sample of \HoB~\cite{Petrenko_2015}.
Although the interpretation of the INS results is not trivial, as the CEF levels show significant dispersion, the measurements performed in the paramagnetic regime at $T = 8$~K revealed the presence of strong excitation lines at 16, 33, 110, and 197~K~\cite{Brunt_thesis_2017}.
Table~\ref{tab:LowESinglets} demonstrates a reasonable agreement between calculated and observed CEF energy levels.
The resulting CEF Stevens parameters $B_{kq}$ are summarized in Table~\ref{tab:CEFParameters}.

With the CEF parameters listed, we have calculated the temperature-dependent magnetic susceptibility $\chi(T)$ and magnetization $M(H)$ for the magnetic field $H$ applied parallel or normal to the $c$~axis.
The calculated curves (not shown) agree reasonably well with the previously published data~\cite{Brunt_2017,Brunt_thesis_2017,Brunt_2018}.
Remarkably, the magnetically easy axis changes direction around 100~K.

\begin{figure}[tb]
\includegraphics[width=0.99\columnwidth]{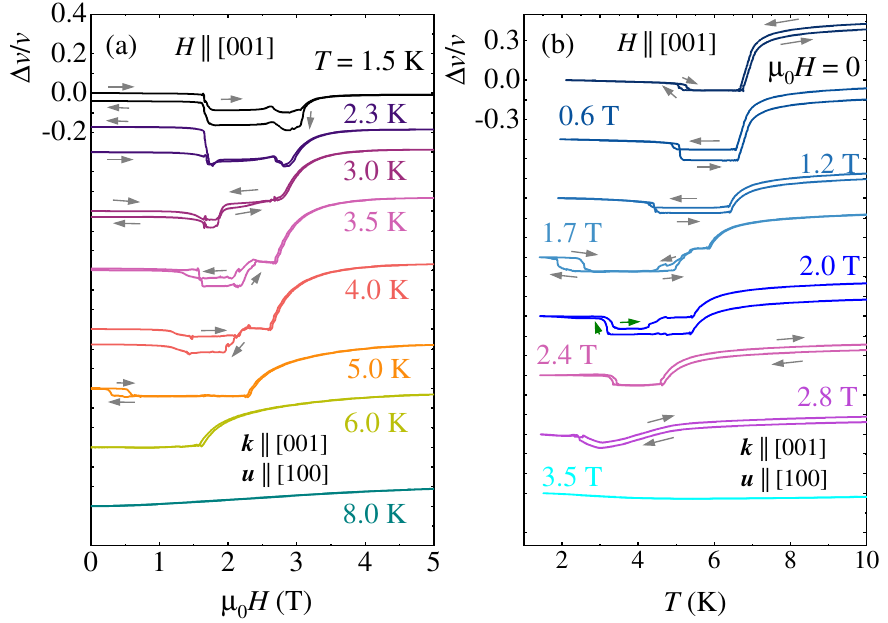}
\caption{The relative sound velocity, $\Delta v/v$, in \HoB\ single crystal as a function of (a) applied magnetic field at constant temperature and (b) temperature in constant applied field.
The gray arrows indicate the decreasing or increasing field or temperature.
The sound velocity curves are consecutively offset for clarity.}
\label{fig:US_in_field}
\end{figure}

We have also collected ultrasound-velocity data in an applied field for $H \parallel c$ (Fig.~\ref{fig:US_in_field}) by either measuring the speed of sound as a function of magnetic field at constant temperature or by varying temperature in a constant field.
In both cases, sharp variations of the speed of sound mark the transitions between different magnetic phases.
The low-temperature magnetic $H-T$ phase diagram for $H \parallel c$ is rather complex; it contains four main (stationary) magnetic phases separated by mixed (transitionary) states~\cite{Brunt_2017}.
The first-order nature of the transitions between various phases is evident from a large hysteresis observed in the speed of sound measurements (Fig.~\ref{fig:US_in_field}), consistent with the previously reported magnetization and neutron diffraction results~\cite{Brunt_2017}.
Although the main feature of the magnetization process in \HoB\ is the formation of an up-up-down ferrimagnetic structure along the $c$~axis leading to a 1/3 magnetization plateau [seen as a sharp reduction in $\Delta v/v$ in the top curve in Fig.~\ref{fig:US_in_field}(a)], detailed knowledge of the transitionary phases surrounding this main phase is still lacking.
The ultrasound data are very useful in locating different states within the phase diagram, but potentially they could also be used to assess the nature of various magnetic phases.

\section{Summary}

From the above results, analysis, and discussion, it is noticeable that the Ho, Dy, and Nd tetraborides all share unusual ground states reflecting the strong spin-orbit-lattice coupling in these systems. 
From detailed low-temperature x-ray diffraction data, we obtain full structural information for \HoB\ across the magnetic phase transitions, providing unambiguous evidence for the monoclinic distortion associated with the quadrupolar ordering.
The transition from tetragonal to monoclinic occurs at $T \approx T_{\rm N2}$.
However, the onset of the lattice instability occurs already upon entering the first antiferromagnetic phase.
This is well above $T_{\rm N2}$ and the ferroquadrupolar ordering, as seen in the large softening of the modulus $C_{44}$.  
The monoclinic symmetry described here is consistent with the symmetry of the magnetic order parameter.
We provide a clear evidence that coupled spin-lattice and quadrupolar terms are key to describe the magnetic ground states in \HoB.

\section{Acknowledgements}
The work at the University of Warwick was supported by EPSRC through grants EP/M028771/1 and EP/T005963/1.
O.A.P. acknowledges the EPSRC grant EP/X020304/1 that sponsored the secondment at the EMFL.
We acknowledge support from the Deutsche Forschungsgemeinschaft (DFG)
through SFB 1143 (Project No.\ 247310070) and the W\"{u}rzburg-Dresden Cluster of Excellence on Complexity and Topology in Quantum Matter--$ct.qmat$ (EXC 2147, Project No.\ 390858490), as well as the support of the HLD at HZDR, member of the European
Magnetic Field Laboratory (EMFL).
We further acknowledge support under the European Union’s Horizon 2020 research and innovation programme through the ISABEL project (No. 871106). 
LTPXRD was carried out at the MGML (mgml.eu), which is supported within the program of Czech Research Infrastructures (project no. LM2023065). This work was also co-funded by the European Union and the Czech Ministry of Education, Youth and Sports (Project TERAFIT - CZ.02.01.01/00/22-008/0004594).
We also acknowledge the Czech Science Foundation (Grant No. 22-22000M) as well as Lumina Quaeruntur fellowship LQ100102201 of the Czech Academy of Sciences.

\bibliography{References_list}
\end{document}